\documentclass{aa} 

\usepackage{psfig}
\usepackage{graphicx}

\authorrunning{Hanasz \& Lesch}
\titlerunning{Fast magnetic reconnection in astrophysical systems}

\begin{document}
\title{Conditions for fast magnetic reconnection \\
in astrophysical plasmas}
\author{Michal Hanasz \inst{1} \and
        Harald Lesch \inst{2}}
 
\institute{
Toru\'n Centre for Astronomy, Nicholas Copernicus University
PL-87148 Piwnice/Toru\'n, Poland, \email{mhanasz@astri.uni.torun.pl}
\and 
University Observatory, M\"unchen University, Scheinerstr. 1, D-81679, Germany and 
Center for Interdisciplinary Plasma Science (CIPS),
\email{lesch@usm.uni-muenchen.de}
}

\date{Received 27 November 2002/ Accepted  11 March 2003}

\setcounter{footnote}{0}

\abstract{ We investigate favourable circumstances for fast magnetic
reconnection in astrophysical plasmas based on recent results by Rogers et al.
(2001).  Given that a critical magnetic field structure with antiparallel field
lines exists, our analysis demonstrates that a sufficient condition for fast
reconnection  is that the ratio of the thermal pressure to the magnetic field
pressure $\beta$  should be  significantly  larger than $ 2 m_e/m_p$ (twice the
ratio of electron mass to proton mass).  Using several  examples
(like the different components of the interstellar medium, the  intergalactic
medium, active galactic nuclei and jets) we show that in almost all
astrophysical plasmas, magnetic reconnection proceeds  fast i.e. independent
of the resistivity, with a few percent of the Alfv{\'e}n speed.  Only for
special cases like neutron stars  and white dwarfs is $\beta$ smaller than $2
m_e/ m_p$. 
}
\maketitle

\section{Introduction}

As is well known, the process of magnetic reconnection is important in many 
space and astrophysical contexts. Since most astrophysical plasmas are 
magnetized, the process of magnetic  reconnection is essential for the
understanding of a broad variety of processes,  like galactic and stellar
dynamos, turbulence, particle acceleration and heating  (e.g Kulsrud 1999;
Lesch 2000). 

Two types  of magnetic reconnection have been identified - slow reconnection
with inflow speeds  significantly lower than the Alfv{\'e}n velocity, and
fast  reconnection which proceeds with an inflow speed close to the Alfv{\'e}n
speed. The latter type of magnetic reconnection is the preferred one since  it
accounts for the fast timescale  associated with solar flares, the solar
magnetic cycle and the topological changes  required for dynamo action in the
interstellar medium of galaxies. The problem  however is to understand how  the
fast reconnection can be realized given the physical  parameters of most
astrophysical plasmas. Since they are highly conducting,  they  are almost
collisionless and can be treated as ideal magnetohydrodynamical  systems in
which magnetic reconnection cannot occur because of zero resistivity. 

The question is what happens when magnetic field lines with antiparallel 
components encounter. This situation is typical for cosmic plasmas, since 
they are agitated by unsaturated external forces like differential  rotation,
winds, explosive motions or turbulence. In such nonequilibrium plasmas, at least
in the initial stages,  the  kinetic energy density is typically much larger
than the magnetic energy  density, i.e. the kinetic pressure is higher than the
magnetic pressure. The  field lines react to these forces by being twisted,
stretched and compressed,  which easily leads to the encounter of antiparallel
field lines, reaching magnetic energy densities  comparable to the kinetic
energy densities of the several external drivers. If  this kind of dynamical
equilibrium is reached, the magnetic field energy is  transferred into current
sheets in which the excess energy is dissipated by  reconnection. In a way,
reconnection is a relaxation mechanism unavoidable  for any plasma which is
externally distorted (Taylor 1986). Thus, magnetic  reconnection is of
fundamental importance for a deeper understanding of  astrophysical plasmas.

The first model to describe such intersecting field lines was developed by 
Parker (1957) and Sweet (1958) in terms of enhanced magnetic diffusion in a
layer  with antiparallel field lines on both sides. Its principle is quite
simple in  terms of mass conservation:  In steady state the magnetic diffusion
velocity  balances the incoming reconnection velocity whereas the plasma inflow
across the sheet is balanced by  plasma outflow along the layer. The plasma
initially entrained on the magnetic  field lines must escape from the
reconnection zone. In the Sweet-Parker scheme  this means a bulk outflow
parallel to the field lines within the layer. The condition  that the plasma
has to leave the reconnection zone is  very important for the effectiveness of
reconnection. The faster plasma expelled is from the layer, the higher 
inflow rate is allowing for a higher reconnection rate.

More quantitatively, starting   from the assumption of stationary Ohmic
dissipation in a three-dimensional  reconnection sheet with an area $\sim L^2$
and a thickness $l$, the dissipation  surface density in the sheet  $lj^2\eta$
acts to reduce the influx of  magnetic energy density $v_r B^2/8\pi$. $v_r$
is the approaching velocity of the  field lines or reconnection speed and
$\eta$ denotes the resistivity.    Including mass conservation, Parker and Sweet
calculated $v_r$ to be of the order  of $v_r\simeq c_A {\rm Rm}^{-1/2}$. ${\rm
Rm}=c_A L/\eta$ denotes the magnetic Reynolds  number which  is a large
quantity in astrophysical plasmas (up to $10^{20}$ for  the interstellar 
medium) and $c_A$ is the Alfv\`en speed. In other words,
Sweet-Parker reconnection is very  slow, too slow for  any reasonable
application in the solar photosphere or even  the Galaxy. 

Later,  Petschek (1964) suggested a model in which shock waves open  up
the outflow channel allowing faster gas outflow and leading to a significantly 
faster inflow of field lines and thus faster reconnection speed. An
X-point-like structure  evolves in which in a localized region magnetic
diffusion is fast. Outside that diffusion layer, shock waves  accelerate the
plasma leading to an open X-point structure. Petschek obtained a  reconnection
speed of  about $c_A/\ln {\rm Rm}$, much faster than the  Sweet-Parker value
and almost independent of the Reynolds number. The Petschek  model is known as
fast reconnection. However, it has been shown (Biskamp 1986, 1996;  Uzdensky
and Kulsrud 2000) that the Petschek solution is not compatible with  uniform or
smooth profiles of the electrical resistivity $\eta$.  

MHD reconnection corresponds to localized current sheets in  which,
due to some resistivity, the energy density $\eta j^2$ is dissipated.  It is
the  value and spatial profile of  electrical resistivity $\eta$ which is the
unknown in astrophysical  plasmas. Normally, $\eta$ is produced by Coulomb
collisions, i.e. $\eta \simeq  T^{-3/2}$. Thus, most astrophysical plasmas are
collisionless, i.e. $\eta \sim 0$ with respect to Coulomb collisions. This is
also the reason why the magnetic Reynolds number ${\rm Rm} \propto1/\eta$ is so
large.  This is the key question for MHD reconnection. 

On the other hand, $\eta$ can be  enhanced due to plasma microinstabilities
which are often excited only in the  reconnection regions where free energy is
available either in the form of a large drift  between ions and electrons or in
strong pressure and magnetic field gradients.  This anomalous resistivity not
only broadens the current sheet thereby increasing the  mass inflow and the
reconnection rate in the context of the Sweet-Parker model  (Kulsrud 2001) but
also its localization is able to open up the outflow channel  for the fast
reconnection ({\bf Sato and Hayashi 1979,} Biskamp and Schwarz 2001). Alternatively,
a recent  theory (Rogers et al. 2001), attempts to explain fast reconnection
rates based on  non-dissipative terms, notably the Hall term in the generalized
Ohm's law. When the physics of  reconnection is associated with the Hall term
as was recently shown by Rogers et al.(2001)  the value of resistivity (if
there is any) is not at all crucial. Nevertheless, any reconnection process
requires that some critical  gradient has been exceeded by the magnetic field
structure to offer the  necessary amount of free energy fed into plasma
fluctuations. Only if such a  critical magnetic field gradient, i.e. a critical
current density, has been  exceeded, magnetic reconnection can start and
proceed. The question is what is the  reconnection speed?

The question of operation  of the fast reconnection in astrophysical  plasmas
is of primary importance. Its presence  would resolve some  problems
related to turbulent dynamos, as well as energetics of stellar coronae,
accretion and galactic disks (Kulsrud \& Anderson 1992, Kulsrud 1999).

It is the aim of this paper to investigate the consequences of 
recent findings in collisionless plasma simulations, namely that magnetic 
reconnection is almost always fast in astrophysical plasmas. In the next
section  we briefly summarize the results of the simulations. Then we 
transform the conditions by Rogers et al.(2001) in order to make them
applicable for astrophysical conditions. Finally, we apply these results  to
astrophysical systems and present some conclusions.

\section{Plasma simulations of collisionless magnetic reconnection}

Recently, several two-fluid and particle simulations have revealed fast rates
of magnetic  reconnection that significantly exceed those of conventional
resistive  magnetohydrodynamic models ((Birn et al. 2001). Such high
reconnection rates  depend sensitively on the formation of an open X-line (as
was already suggested  by Petschek (1964)), i.e. the thickness of the
reconnection layer has to increase with  distance from the X-point ( Shay et
al. 1998). It is the small-scale dynamics that provide the fast  reconnection
dynamics. Small scales mean the electron skin depth $d_e=c/\omega_{pe}$ and ion
skin depth $d_i=c/\omega_{pi}$, respectively. Here $\omega_{pe}$ denotes the 
electron plasma frequency and $\omega_ {pi}$ is the ion plasma frequency. 
Furthermore, at these small scales the electrons decouple from the ions. 
Electrons are strongly magnetized and their flow scales inversely with the
width  of the layer. When the layer shrinks the electrons are accelerated. This
leads to  an electron flux from the layer which remains large although the
reconnection  layer size decreases. Of course the ions follow and the plasma is
expelled  with high speeds from the reconnection region. The reconnection rate
becomes insensitive to the mechanism  that is responsible for the nonidealness
(Rogers et al. 2001).

\begin{figure}
 \centering
  \resizebox{\hsize}{!}{\includegraphics{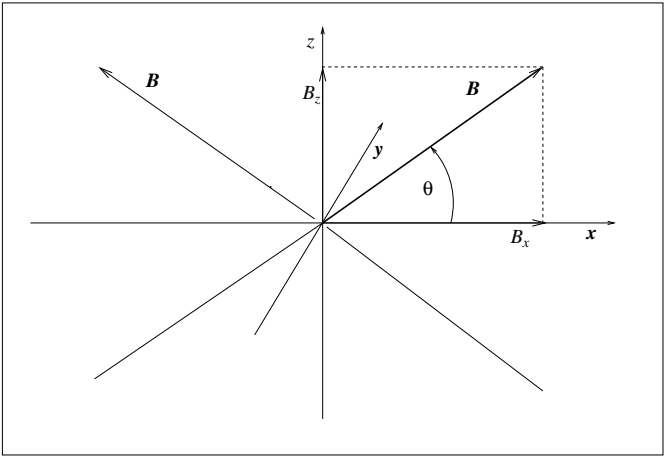}}
  \caption{Geometry of the magnetic field in the vicinity of the current sheet 
  before the onset of reconnection. The current sheet corresponding to the
  contact discontinuity of magnetic field is coplanar with the $xz$ plane.
  The two magnetic field vectors inclined with respect to the $x$-axis 
  represent two magnetic field lines on both sides of the current sheet.}
  \label{field-geom}
\end{figure}

In Fig.~\ref{field-geom} we illustrate the 3-D geometry of the magnetic field
in Cartesian coordinates in the vicinity of the current sheet.   Before the
onset of reconnection  the sheet is coplanar with the $xz$-plane. The $y$-direction
is perpendicular to the current sheet. 

The plasma $\beta$ parameter is related to the total magnetic field
$\boldmath{B}$ and the angle $\theta$ denotes the inclination of $\boldmath{B}$
with respect to the $x$-axis. $B_x = B \cos \theta$ is the reconnecting
component of the magnetic field and $B_z = B \sin \theta$ is the guiding field.
The $B_y$ component is zero prior to the reconnection event. It becomes
finite as a result of reconnection.

The dynamics of the reconnection layer have been reproduced by simulations of 
Rogers et al. (2001) that show that the combined action of whistler waves and
kinetic Alfv{\'e}n waves  play the central role in producing the open outflow
region what characterizes  the two-fluid and particle simulations. Both waves
obey dispersion relations  $\omega\propto k^2$, i.e. their phase velocities
$v_\phi\propto k$ increase with  decreasing spatial scale. 
For whistler waves
the dispersion relation is $\omega  = k^2 c_{Ak} d_i$.
Kinetic Alfv{\'e}n waves obey the relation  $\omega=k r_{gi}k_\parallel c_A$,
where $k_\parallel$ denotes the wave vector  parallel to the magnetic field. 

Rogers et al. (2001) show that the dynamics of reconnection  is
related to the presence of  whistler and kinetic Alfv{\'e}n waves that is
controlled by two dimensionless parameters: 
\begin{eqnarray}
\frac{\beta_k}{2}& =& \frac{c_s^2}{c_{Ak}^2} \label{W} \\
\mu_k            & =& \frac{c_m^2}{c_{Ak}^2} \frac{m_e}{m_i} \label{KA}
\end{eqnarray}
where $c_s^2 = k_B(T_e+T_i)/m_i$, 
$c_{Ak}^2=(\vec{B}\cdot\vec{k}/k)^2/(4\pi\rho)$,
$c_m^2=c_A^2+c_s^2$ and $c_A^2 = B^2/(4\pi\rho)$.
$\beta_k$ and $\mu_k$ measure the strength of the guiding magnetic field and the
plasma pressure.  The two-parameter space $(\beta_k,\mu_k)$ splits into  four
regimes (Fig.~2 in Rogers et al. 2001) admitting different combinations of
whistler and kinetic Alfv{\'e}n waves:

\begin{enumerate}
\item $\mu_k \ll 1$, $\beta_k/2 \gg 1$ - both whistler and kinetic Alfv{\'e}n waves
       possible.
\item $\mu_k \ll 1$, $\beta_k/2 \leq 1$ - only whistler waves possible

\item $\mu_k \geq 1$, $\beta_k/2 \gg \mu_k$ - only kinetic Alfv{\'e}n waves
       possible.
\item $\mu_k \geq 1$, $\beta_k/2 \leq \mu_k$ - no quadratic dispersive waves.
\end{enumerate}

The numerical simulations presented by Rogers et al. (2001) demonstrate that
the condition of fast magnetic reconnection of the Petschek type is related
to the existence of dispersive waves of the two types mentioned above, i.e.
fast magnetic reconnection operates in the first three regimes.  On the other
hand, only the absence of dispersive waves in regime 4 allows for the slow
Parker-Sweet model.

The results of numerical simulations are interpreted in terms of wave analysis.
The fast reconnection is possible when the X-type pattern of separatrices is
stationary. Its existence is equivalent to the presence of a stationary
magnetic field component $B_y$ perpendicular to the reconnecting component
$B_{x0}$. The magnitude of the $B_y$ component, which is limited to $B_{x0}$, is
not known {\em a priori}  since there is no  $B_y$ in the initial state and
later on $B_y$ appears as a result of reconnection. 

The transition from the initial state (before the onset of reconnection) to
the appearance of a finite  $B_y$ component has to be time-dependent, although
this dependence is unknown. Rogers et al. (2001) discuss conditions for the
mentioned two types of dispersive waves propagating in the $y$ direction
($\boldmath{k} = k \hat{e}_y$). They consider first the parameters  $\beta_y$,
and $\mu_y$, both depending on  $B_y$, thus unknown functions of time.  They
find a relation between $\mu_y$ and $\beta_y/2$. $B_x$ is vanishing 
at the $xz$-plane for $y=0$, thus 
\begin{equation}
\mu_y = \frac{c_{Ay}^2+c_{Az}^2 + c_s^2}{c_{Ay}^2}\frac{m_e}{m_i} 
\end{equation}
can be expressed as
\begin{equation}
\mu_y = \frac{m_e}{m_i} + C \frac{\beta_y}{2} \simeq C \frac{\beta_y}{2},
\label{mubeta}
\end{equation}
where $C = m_e/m_i (1+2/\beta_z)$ and $\beta_z=c_s^2/c_{Az}^2$ is the plasma
$\beta$ associated with the guiding field $B_z$. Eq.~(\ref{mubeta}) describes a
trajectory of the reconnection process in the $(\mu_y, \beta_y)$-parameter
space for an unknown time dependence of the trajectory. It turns out,
according to conditions 1.-4., that if $C < 1$, the whole trajectory lies in
the regimes 1, 2 and 3 where the fast reconnection operates. 
The constant  $C$ parametrizes the magnitude of the guiding field.
The condition $C < 1$ is equivalent to the statement that the guiding field is
below a certain critical value corresponding to  $\beta_z = 2 m_e/m_i$,
and  $C > 1$ means that the critical guiding field is exceeded.
In the latter case the trajectory lies only partially in region 2
corresponding to whistler waves ($\mu_y < 1$). The remaining part of the
trajectory is placed in region 4 which does not admit dispersive waves. 
Therefore the condition for fast reconnection $\mu_y < 1$    
implies that
\begin{equation}
\frac{\beta_y}{2} \simeq \frac{1}{C}, 
\end{equation}
i.e. a certain threshold of $B_y$ has to be exceeded.  
However, the vanishing $B_y$ at $t=0$ places the initial configuration
just in region 4.

Therefore $\beta_z$ should not be too small to initiate the fast reconnection, 
\begin{equation}
\beta_z \geq \frac{2 m_e}{m_i}\label{maxbz} ,
\end{equation}
which is equivalent to
the statement that the guiding field should not be too strong.

The condition (\ref{maxbz}) is a sufficient,  but not necessary, condition
for fast reconnection.  This means that if condition (\ref{maxbz}) is  met, fast
reconnection will occur.  But this does not mean that  fast reconnection will
not occur if condition (\ref{maxbz}) is not met.  In  particular, large
fluctuations or small-scale instabilities (which  typically occur in a narrow
current layer) could lead to a large  enough $B_y$ so as to initiate fast
reconnection.

Although the last mentioned effects provide an opportunity for fast
reconnection  in the regime $\beta_z < 2 m_e/m_i$, we cannot quantify them on
theoretical ground nor detect them observationally in astrophysical
systems. For that reason we treat the condition (\ref{maxbz}) as a safe
limitation of the fast reconnection process in the parameter space.  

The conditions (\ref{W}) and (\ref{KA}) for whistler and kinetic Alfv{\'e}n
wave dynamics can be expressed in terms of parameters describing of the medium
far away from the current sheet or parameters before the onset of reconnection.

Rogers et al. (2001) conclude that: ''The condition for whistler dynamics to be
present,

\begin{equation}
B_x^2 > B^2\left(1+\frac{\beta}{2}\right)\frac{m_e}{m_i}, \label{minbx}
\end{equation}
is satisfied in many system of physical interest. Assuming this condition is
satisfied, the physics at the smallest scales $\sim c/\omega_{pe}$
characteristic for the dissipation region are always governed by whistler
waves''. On the other hand ``if plasma beta is sufficiently high

\begin{equation}
\frac{\beta}{2} \geq \max \left[\left(1+\frac{\beta}{2}\right)\frac{m_e}{m_i},
                                    \frac{B_x^2}{B^2}\right],\label{minbeta}
\end{equation}
then kinetic Alfv{\'e}n dynamics also play a role at somewhat larger scales''.

\section{The conditions for fast reconnection in randomly distorted
astrophysical plasmas}

In order to apply the findings by Rogers et al. (2001) to astrophysical plasmas
we note that with the exception of  the earth magnetosphere, where in-situ
satellite measurements are possible, the relevant spatial scales of individual
reconnection events are much too  small to be spatially resolved  by any
astrophysical observation. This means that an evaluation of the conditions for
fast reconnection  is not possible for individual current sheets in distant
astrophysical objects. The only way is to investigate the parameter space
$(\beta_k,\mu_k)$ in different variables $(\beta,\theta)$, where $\beta$ is
observable and $\theta$ is an individual property of each current sheet. Since
$\theta$ is not available from observations, the evaluation of conditions for 
fast reconnection in astrophysical objects is not possible, unless we
incorporate additional knowledge about the systems under consideration.

This necessary additional piece of information has to rely on the fact that
even in cases of very strong magnetic fields a full range of angles $\theta$ can
be expected (in a statistical sense) due to the fact that a huge reservoir of
gravitational potential energy is available for random distortions of magnetic
field lines. Therefore $\theta$ can be treated as a random variable. In such a
case the determination whether fast reconnection is possible or not depends on 
the  existence of a range of angles $\theta$ admitting reconnection. If such a
range for given $\beta$ exists, fast reconnection is possible. Thus
the two dimensional parameter space can be reduced to a one dimensional space
of plasma-$\beta$. In the following considerations we perform such a
transformation of the parameter space. 

The conditions (\ref{minbx}) and (\ref{minbeta}) can be expressed in terms of
two parameters: plasma-$\beta$ and the pitch angle $\theta$. The conditions for
whistler wave dynamics and kinetic Alfv{\'e}n wave dynamics are 
\begin{equation}
\left(1+\frac{\beta}{2}\right)\frac{m_e}{m_i} < \cos^2 \theta
\end{equation}
and
\begin{equation}
\frac{\beta}{2} > \max \left[\left(1+\frac{\beta}{2}\right)\frac{m_e}{m_i},
                  \cos^2 \theta \right],
\end{equation}
respectively.
 
The condition (\ref{maxbz}) for an upper limit for the magnitude of the guiding
magnetic field can be analogously written in the form

\begin{equation}
\frac{\beta}{2} > \frac{m_e}{m_i}\sin^2 \theta .\label{maxbz1}
\end{equation} 

\begin{figure}  
  \resizebox{\hsize}{!}{\includegraphics{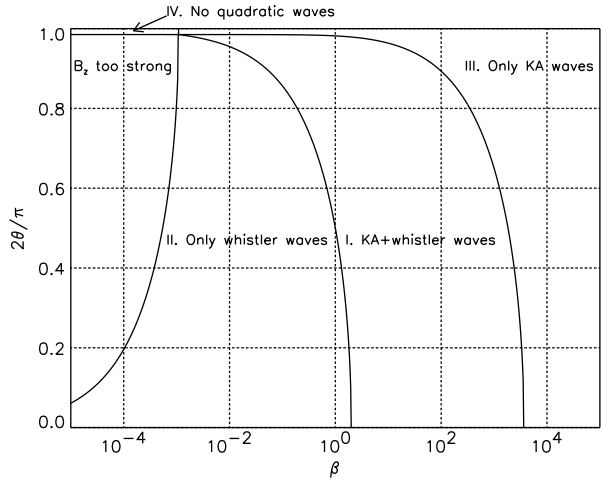}}
\caption{Parameter space for quadratic waves. Numbering of different regions
follows the list following conditions (\ref{W}) and (\ref{KA}). An additional
region on the left of the figure results form the upper limit on the guiding
field $B_z$.}
\label{param-space}  
\end{figure}

The conditions above for dispersive wave dynamic expressed in terms of
parameters $(\beta,\theta)$ are presented in Fig.~\ref{param-space}.

The division of the parameter space shown in Fig.~\ref{param-space} is
basically the same as displayed in Fig.~2 of Rogers et al. (2001). The only
difference is that we divided region II (admitting only whistler waves) into
two subregions following the condition (\ref{maxbz1}) resulting from the
upper limit of the guiding magnetic field.

The aim of the modified parametrization of the results by Rogers et al. (2001)
is to determine the range of pitch angles $\theta$ admitting fast reconnection
for a given value of plasma-$\beta$. We will subsequently implement the new
formulation of the conditions for fast reconnection to a turbulent medium.

Keeping in mind the fact that we consider small regions of a medium disturbed
by  external forces we can assume that the distribution of pitch angles
$\theta$ is uniform in the range $[0,\pi/2]$. Therefore, one can expect fast
reconnection for a given value of plasma-$\beta$ even if a narrow range of
pitch angles fulfills the conditions for fast reconnection. We note however
that efficiency of the reconnection will be dependent on the widths of the
range of appropriate pitch angles.

We note that both kinds of dispersive waves operate for a wide range of pitch
angles (excluding cases of a very weak reconnecting component $B_x$) for
plasma-$\beta$ between one and several $10^3$. This range is extremely
important for astrophysical plasmas. For $\beta$ above several $10^3$ fast
reconnection is associated with kinetic Alfv{\'e}n waves.  For $\beta$ varying
in between $2m_e/m_i$ and 1 the whistler wave dynamics contributes to fast
reconnection and moreover there is a range of large pitch angles admitting
operation of both kinds of dispersive waves.  

Considering the lowest values of plasma-$\beta$ up to $2m_e/m_i$ we note that
only a narrow range of pitch angles (region IV) does not admit quadratic waves,
however the adjacent region of smaller pitch angles 
requires large fluctuations or other types of small scale instabilities
 to exceed a threshold in $B_y$ for strong
guiding fields $B_z$. 
Fast reconnection is still possible due to whistler waves at very low
plasma-$\beta$ only if the field lines on opposite sides of the current sheet
are almost antiparallel.

Now we shall discuss astrophysical consequences of the above findings. The most
common astrophysical circumstance is a turbulent medium that is agitated by
external forces and/or instabilities. The natural behaviour of relaxing MHD
systems is the  spontaneous formation of current sheets.  The encountering
regions with non-parallel magnetic fields are filled with magnetic fields of
typical strengths which is, in principle, an observable quantity. On the other
hand the geometry of field lines around the contact discontinuity is usually
unknown. Therefore, only the typical magnitude of plasma $\beta$ can be
estimated for a particular system. 

Thus, we can conclude that the conditions for fast magnetic reconnection in a
turbulent medium can be related to the typical magnitude of plasma $\beta$ in
the  medium considered.

\section{Dynamics of astrophysical plasmas}

\begin{table*}
\begin{tabular}{lcccc} 
\hline
\hline
Type of medium      & Density & Mag. field  & Temp & Plasma $\beta$  \\
Unit        &cm$^{-3}$&   & K   &   \\
\hline
Gas in supercusters of gal. &$10^{-6}$& 0.5$\mu$G &$10^7$ & 0.14   \\
Gas in clusters of gal.   &$10^{-4}$& 1$\mu$G &$10^7$ & 3.5  \\
Gas in galactic halos   &$10^{-3}$& 4$\mu$G &$10^6$ &0.22 \\
Spiral arm interstellar gas.  &0.5  &   4$\mu$G &$10^4$ &1.09 \\
Large interstellar shells-initial&0.8 &   4$\mu$G &$10^3$ &0.3  \\
HI gas in diffuse clouds  & 3 &   5$\mu$G & $100$ &0.04 \\
HII regions       & 5 &  10$\mu$G &$10^4$ &1.74 \\
HI gas in interclumps   & 90  &  15$\mu$G &$100$  &0.14 \\
HI gas in abs. initial    &100  &  15$\mu$G & 10  &0.02 \\
Solar photosphere (spots) &$10^{17}$&2$\times 10^3$G& $10^4$&0.8  \\
Solar corona      &$10^8$ & $10$G   &$10^6$&$3.5\times10^{-3}$  \\
Magnetic stars      &$10^{10}$& $10^3-10^4$G  &$10^6$&$3\times10^{-5}$  \\
Bipolar flows     &$10^3$ &  $10^{-4}$G    & $10$&$3\times10^{-3}$\\
AGN nuclei      &$10^{10}$&$10^{-3}-10^{-1}$G& $10^6-10^8$&0.3  \\
AGN jets    &$10^{-4}$&$10^{-5}-10^{-3}$G&$10^6-10^7$&$10^{-2}$ \\
Neutron star surface    &$10^{12}$&$10^{12}$G& $2\times10^{6}$&$10^{-20}$ \\
Magnetized white dwarfs      &$10^{12}$& $10^6$G  &$10^6$&$10^{-8}$  \\
\hline
\end{tabular} 
\caption{Parameters and plasma $\beta$ for various astrophysical objects} 
\end{table*}

In Table 1 (partially from Vall{\'e}e 1995 and from Tajima \& Shibata 1997) we
present  the most  important plasma parameters: particle density, magnetic field
strength and  temperature for several astrophysical object classes. With these
values we can estimate the value of $\beta =8\pi nk_B T/B^2$. Obviously, only 
for neutron stars, magnetized white dwarfs and in magnetic stars  the
ratio of thermal energy density to magnetic energy density is significantly 
smaller than one.  For plasmas like the ionized gas in superclusters, in galaxy
clusters, in galactic halos, in spiral arms, the neutral HI gas in interclumps
and the ionized gas in HII  regions, the plasma beta is of the order of one or
somewhat larger.

Combining these numbers with the results from completely different 
reconnection simulations, we can conclude that in almost every astrophysical 
plasma, magnetic reconnection proceeds at a fast rate.

What is the physical reason for the distribution of $\beta $?  The answer is
that
the magnetic fields in cosmic plasmas are the result of some  processes which
convert kinetic plasma energy into magnetic energy. The kinetic  energy of a
cosmic plasma is the result of the different forces acting on the ionized gas.

Of course, the most important force in the universe is gravity, but in many
systems like  stellar accretion disks, disks around black holes, planetary
systems and disk  galaxies, it is rotation which at least temporarily  balances
gravity.  This balance results in differentially rotating  plasmas, i.e. shear
flows.  Other sources of shear flows are stellar winds, stellar explosions
and jets. In molecular clouds it is a mixture of turbulent motions and shear
flows that  dominate the plasma dynamics. In any case, it is the kinetic
energy density  $1/2\rho v^2$ which represents the ultimate source of the
magnetic field energy. The velocity $v$  may be due to turbulence, rotation or
a directed large-scale flow. Such external  unsaturated forces like
differential rotation, gravity and/or explosions, winds  and jets agitate the
magnetized plasmas. That is, the amplification of magnetic  fields is
equivalent to a transfer of plasma kinetic energy into magnetic  energy.
Therefore, magnetic fields can hardly grow to field strengths larger  than
equipartition value given by $1/2\rho v^2=B^2/8\pi$. In a virialized plasma we
can reasonably expect  equipartition between kinetic energy density and and
thermal energy density $n  k_B T$, as well. In non steady plasmas, kinetic
energy will dominate over the  magnetic energy density. 

There are two possibilities for deviations from these rather general arguments: 

1. The frozen-in magnetic field is under the influence of gravitational
collapse of a stellar  object, like in the case of white dwarfs and neutron
stars. There the plasma  pressure is negligible compared to the gravitational
potential, i.e. the magnetic field  strength is increasing as the size of the
system shrinks due to gravitational  collapse which is stopped by the pressure
of the degenerated stellar material.  Since stars are spheres, they prefer
dipolar magnetic field  configurations that connect the star with its
environment by a magnetosphere.  When the rigidly rotating magnetospheres reach
the Alfv{\'e}n speed the magnetic  field decouples from the stellar field and
winds up into a solar wind-like  spiral configuration which is at some distance
in equipartition with the  pressures in the remnant region.

2. Only for small spatial scales the magnetic field energy density can be
significantly larger than the thermal energy density. That happens in stellar
coronae and photospheres. There, the  gas density drops faster than the
magnetic field strengths. Due to footpoint motions of the  stellar surface the
magnetic field lines that escape into the photospheres or  coronae represent
an energy density that may be higher than the thermal energy  density. But as
we know from the Sun, even in the solar corona the plasma beta  is not as small
as that required to make reconnection proceed at a slow rate. On  the contrary,
the fine structure and the energy requirements of solar flares were some of the
major arguments  supporting fast reconnection. To understand very rapid flares,
reconnection must  be very fast. In the solar corona
the plasma $\beta$ is small but still, fast reconnection is at work. 
This supports for the recent  findings of plasma simulation groups.

\section{Conclusions}

We have applied the results of the most sophisticated up to date simulations 
of magnetic reconnection to astrophysical plasmas. There is a long-standing 
discussion as to whether reconnection proceeds slowly (according to the 
Sweet-Parker mechanism) with a velocity significantly smaller than the
Alfv{\'e}n  speed, or fast with a few percent of the Alfv{\'e}n speed. A
definite answer to this  problem is very important to understand dynamo action in stars
and galaxies, as well as  for the understanding of stellar flares. All these
phenomena require the action  of fast reconnection. 

First, we emphasize that magnetic reconnection can only proceed if field lines 
with antiparallel directions are close enough. In other words, only if some
critical value for  the current density has been exceeded is enough free energy is
available for  the excitation of plasma fluctuations necessary to induce
enhanced resistivity. In other words, fast reconnection can start after the
current sheets appear. However, formation of current sheets may introduce an
additional time delay for the onset of reconnection. This time delay should be
rather attributed to macroscopic properties of the system and may be
responsible for the criticality of the reconnection phenomena even in the case of
plasma $\beta$ of the order of one.

Recently a number of plasma simulations demonstrated that magnetic
reconnection should be fast in almost every circumstance provided the critical 
field structure has been established by external forces acting on the magnetic 
field lines.
Especially the results of Rogers et al.  (2001) lead to the conclusion that 
reconnection is unconditionally fast if the magnetic field  strength is not too
high i.e. if  $\beta \geq 2 m_e/m_p$. For $\beta < 2 m_e/m_p$, fast reconnection
is conditionally possible except in a narrow range of pitch angles close to $\pi/2$
(region IV of small reconnecting magnetic field component) if  large
fluctuations or small-scale instabilities could lead to a large  enough $B_y$
or the pitch angle $\theta$ is sufficiently small (near antiparallel magnetic
fields on both sides of the current sheet). 

If the field is not so strong, reconnection is fast  and its rate does not
depend on the resistivity. Lazarian and Vishniac (2000) already  concluded from
their investigations of turbulent magnetic reconnection that this  process
becomes fast when field stochasticity is taken into account. As a  consequence
solar and galactic dynamos are also fast, i.e. do not depend on fluid 
resistivity. We support their conclusions by checking the beta-values for 
different astrophysical systems. Only for the very strongly magnetized
stellar  remnants (neutron stars and magnetized white dwarfs) the possibility
of fast reconnection depends on additional conditions listed above.

 In stellar photospheres, where the beta-values are also small but larger than
$2 m_e/m_p$, the magnetic  field  is still not rigid enough to inhibit fast
reconnection, as is proven by  the fast solar flares. For all other cosmical
plasmas, reconnection will proceed  at a rate comparable to the Alfv{\'e}n
speed.

From our investigation we can conclude that especially in weak magnetic fields 
reconnection can proceed at a fast rate allowing for fast dynamo action also
in young galaxies.

\begin{acknowledgements}   We thank the referee Dr. James F. Drake for helpful
comments. This work was supported by Polish Committee for  Scientific Research
(KBN) through the grant PB 404/P03/2001/20.  The presented work is a continuation
of a research program realized by  MH under the financial support of {\em
Alexander von Humboldt Foundation}.   \end{acknowledgements}  

\bibliographystyle{unsrt}

\end{document}